\pdfoutput=1
\documentclass[a4paper,11pt, english]{article} 
\usepackage{pool/jheppub}
\usepackage{babel}
\usepackage{graphbox}
\usepackage{multirow}
\usepackage{makecell}
\usepackage{amsmath,amssymb, gensymb}
\usepackage{natbib,hyperref}
\usepackage{url}
\usepackage{siunitx}
\usepackage{xspace}
\usepackage{microtype}
\usepackage{subfig}
\usepackage[useregional]{datetime2}
\usepackage{graphicx}
\usepackage{textcomp}

\usepackage{siunitx}
\usepackage{xspace}

\newcommand{\NEW}{NEXT-White}

\newcommand{\Fig}{Figure}

\newcommand{\bbonu}{\ensuremath{\beta\beta0\nu}}

\newcommand{\Qbb}{\ensuremath{Q_{\beta\beta}}}

\newcommand{\CS}{\ensuremath{^{137}}Cs}

\newcommand{\TL}{\ensuremath{{}^{208}\rm{Tl}}}

\newcommand{\THO}{\ensuremath{{}^{232}{\rm Th}}}
\newcommand{\TO}{\ensuremath{^{228}}Th}

\newcommand{\Kr}[1]{\ensuremath{^{#1}\mathrm{Kr}}\xspace}

\newcommand{\Tl}[1]{\ensuremath{^{#1}\mathrm{Tl}}\xspace}

\DeclareSIUnit\c{\mbox{$c$}}
\DeclareSIUnit\magn{\mbox{$\times$}}
\DeclareSIUnit\min{min}
\DeclareSIUnit\week{week}
\DeclareSIUnit\year{yr}
\DeclareSIUnit\years{years}
\DeclareSIUnit\yr{yr}
\DeclareSIUnit\standard{std}
\DeclareSIUnit\str{sr}
\DeclareSIUnit\ppm{ppm}
\DeclareSIUnit\ppb{ppb}
\DeclareSIUnit\ppt{ppt}
\DeclareSIUnit\pe{PE}
\DeclareSIUnit\spe{SPE}
\DeclareSIUnit\ev{events}
\DeclareSIUnit\ct{counts}
\DeclareSIUnit\neutron{\mbox{$n$}}
\DeclareSIUnit\smp{samples}
\DeclareSIUnit\Sample{S}
\DeclareSIUnit\ch{ch}
\DeclareSIUnit\hit{hit}
\DeclareSIUnit\hits{hits}
\DeclareSIUnit\bin{(\mbox{5-PE}~bin)}
\DeclareSIUnit\sgm{\mbox{$\sigma$}}
\DeclareSIUnit\rms{RMS}
\DeclareSIUnit\keVr{\mbox{keV$_{\rm nr}$}}
\DeclareSIUnit\keVee{\mbox{keV$_{e{\rm e}}$}}
\DeclareSIUnit\ph{photon}
\DeclareSIUnit\pes{pes}
\DeclareSIUnit\el{electrons}
\DeclareSIUnit\pm{PMT}
\DeclareSIUnit\inch{"}
\DeclareSIUnit\bit{bit}
\DeclareSIUnit\sample{samples}
\DeclareSIUnit\barn{barn}
\DeclareSIUnit\bara{bar}
\DeclareSIUnit\barg{barg}
\DeclareSIUnit\mlardepth{\mbox(meter~of~\LAr~depth)}
\DeclareSIUnit\Curie{Ci}
\DeclareSIUnit\psi{psi}
\DeclareSIUnit\parsec{pc}
\DeclareSIUnit\liveday{\mbox{live-days}}
\DeclareSIUnit\days{\mbox{days}}
\DeclareSIUnit\day{\mbox{day}}
\DeclareSIUnit\miles{\mbox{miles}}
\DeclareSIUnit\degreeC{\mbox{$^{\circ}$C}}
\DeclareSIUnit\electron{\mbox{$e^-$}}
\DeclareSIUnit\Euro{\mbox{\euro}}
\DeclareSIUnit\cph{cph}
\DeclareSIUnit\neq{neq}
\DeclareSIUnit\unit{unit}
\DeclareSIUnit\byte{Byte}
\DeclareSIUnit\Bq{\becquerel}

\newcommand{\KrEnergy}{\SI{41.5}{\keV}}

\newcommand{\CsGammaEnergy}{\SI{661.6}{\keV}}

\newcommand{\NewSevenBarPressureRunII}{\SI{7.2}{\bar}}

\newcommand{\NewPressureVesselMaterial}{316Ti}

\newcommand{\NewTpcLength}{\SI{664.5}{\mm}}

\newcommand{\NewTpcDriftLength}{\SI{530.3 +- 2}{\mm}}

\newcommand{\NewTpcELGap}{\SI{6}{\mm}}

\newcommand{\NewNumberOfSiPM}{\num{1792}}

\newcommand{\NewSipmPitch}{\SI{10}{\mm}}

\newcommand{\NewNumberOfPMT}{\num{12}}

\newcommand{\NewCathodeToPMTs}{\SI{130}{\mm}}

\newcommand{\NewTpcDiameter}{\SI{454}{\mm}}

\newcommand{\NewTypePMT}{Hamamatsu R11410-10}

\newcommand{\NewBarrelICS}{\SI{60}{\mm}}

\newcommand{\NewPlatesICS}{\SI{120}{\mm}}

\begin{document}
\title{Energy calibration of the \NEW\ detector with 1\% resolution near Q$_{\beta\beta}$ of $^{136}$Xe}

\collaboration{The NEXT Collaboration}
\author[18,a]{J.~Renner,\note[a]{Corresponding author.}}
\author[20,15]{G.~D\'iaz~L\'{o}pez,}
\author[15,9]{P.~Ferrario,}
\author[20]{J.A.~Hernando~Morata,}
\author[18]{M.~Kekic,}
\author[18,20,b]{G.~Mart\'inez-Lema,\note[b]{Now at Weizmann Institute of Science, Israel.}}
\author[15]{F.~Monrabal,}
\author[15,9,c]{J.J.~G\'omez-Cadenas,\note[c]{NEXT Co-spokesperson.}}
\author[2]{C.~Adams,}
\author[18]{V.~\'Alvarez,}
\author[6]{L.~Arazi,}
\author[19]{I.J.~Arnquist,}
\author[4]{C.D.R~Azevedo,}
\author[2]{K.~Bailey,}
\author[21]{F.~Ballester,}
\author[18]{J.M.~Benlloch-Rodr\'{i}guez,}
\author[13]{F.I.G.M.~Borges,}
\author[3]{N.~Byrnes,}
\author[18]{S.~C\'arcel,}
\author[18]{J.V.~Carri\'on,}
\author[22]{S.~Cebri\'an,}
\author[19]{E.~Church,}
\author[13]{C.A.N.~Conde,}
\author[11]{T.~Contreras,}
\author[18]{J.~D\'iaz,}
\author[5]{M.~Diesburg,}
\author[13]{J.~Escada,}
\author[21]{R.~Esteve,}
\author[18]{R.~Felkai,}
\author[12]{A.F.M.~Fernandes,}
\author[12]{L.M.P.~Fernandes,}
\author[4]{A.L.~Ferreira,}
\author[12]{E.D.C.~Freitas,}
\author[15]{J.~Generowicz,}
\author[11]{S.~Ghosh,}
\author[8]{A.~Goldschmidt,}
\author[20]{D.~Gonz\'alez-D\'iaz,}
\author[11]{R.~Guenette,}
\author[10]{R.M.~Guti\'errez,}
\author[11]{J.~Haefner,}
\author[2]{K.~Hafidi,}
\author[1]{J.~Hauptman,}
\author[12]{C.A.O.~Henriques,}
\author[15,18]{P.~Herrero,}
\author[21]{V.~Herrero,}
\author[6,7]{Y.~Ifergan,}
\author[2]{S.~Johnston,}
\author[3]{B.J.P.~Jones,}
\author[17]{L.~Labarga,}
\author[3]{A.~Laing,}
\author[5]{P.~Lebrun,}
\author[18]{N.~L\'opez-March,}
\author[10]{M.~Losada,}
\author[12]{R.D.P.~Mano,}
\author[11]{J.~Mart\'in-Albo,}
\author[15]{A.~Mart\'inez,}
\author[3]{A.D.~McDonald,}
\author[12]{C.M.B.~Monteiro,}
\author[21]{F.J.~Mora,}
\author[18]{J.~Mu\~noz Vidal,}
\author[18]{P.~Novella,}
\author[3,d]{D.R.~Nygren,\note[d]{NEXT Co-spokesperson.}}
\author[18]{B.~Palmeiro,}
\author[5]{A.~Para,}
\author[23]{J.~P\'erez,}
\author[3]{F.~Psihas,}
\author[18]{M.~Querol,}
\author[2]{J.~Repond,}
\author[2]{S.~Riordan,}
\author[16]{L.~Ripoll,}
\author[10]{Y.~Rodr\'iguez Garc\'ia,}
\author[21]{J.~Rodr\'iguez,}
\author[3]{L.~Rogers,}
\author[15,23]{B.~Romeo,}
\author[18]{C.~Romo-Luque,}
\author[13]{F.P.~Santos,}
\author[12]{J.M.F. dos~Santos,}
\author[6]{A.~Sim\'on,}
\author[14,f]{C.~Sofka,\note[f]{Now at University of Texas at Austin, USA.}}
\author[18]{M.~Sorel,}
\author[14]{T.~Stiegler,}
\author[21]{J.F.~Toledo,}
\author[15]{J.~Torrent,}
\author[18]{A.~Us\'on,}
\author[4]{J.F.C.A.~Veloso,}
\author[14]{R.~Webb,}
\author[6,g]{R.~Weiss-Babai,\note[g]{On leave from Soreq Nuclear Research Center, Yavneh, Israel.}}
\author[14,h]{J.T.~White,\note[h]{Deceased.}}
\author[3]{K.~Woodruff,}
\author[18]{N.~Yahlali}
\emailAdd{josren@uv.es}
\affiliation[1]{
	Department of Physics and Astronomy, Iowa State University, 12 Physics Hall, Ames, IA 50011-3160, USA}
\affiliation[2]{
	Argonne National Laboratory, Argonne, IL 60439, USA}
\affiliation[3]{
	Department of Physics, University of Texas at Arlington, Arlington, TX 76019, USA}
\affiliation[4]{
	Institute of Nanostructures, Nanomodelling and Nanofabrication (i3N), Universidade de Aveiro, Campus de Santiago, Aveiro, 3810-193, Portugal}
\affiliation[5]{
	Fermi National Accelerator Laboratory, Batavia, IL 60510, USA}
\affiliation[6]{
	Nuclear Engineering Unit, Faculty of Engineering Sciences, Ben-Gurion University of the Negev, P.O.B. 653, Beer-Sheva, 8410501, Israel}
\affiliation[7]{
	Nuclear Research Center Negev, Beer-Sheva, 84190, Israel}
\affiliation[8]{
	Lawrence Berkeley National Laboratory (LBNL), 1 Cyclotron Road, Berkeley, CA 94720, USA}
\affiliation[9]{
	Ikerbasque, Basque Foundation for Science, Bilbao, E-48013, Spain}
\affiliation[10]{
	Centro de Investigaci\'on en Ciencias B\'asicas y Aplicadas, Universidad Antonio Nari\~no, Sede Circunvalar, Carretera 3 Este No.\ 47 A-15, Bogot\'a, Colombia}
\affiliation[11]{
	Department of Physics, Harvard University, Cambridge, MA 02138, USA}
\affiliation[12]{
	LIBPhys, Physics Department, University of Coimbra, Rua Larga, Coimbra, 3004-516, Portugal}
\affiliation[13]{
	LIP, Department of Physics, University of Coimbra, Coimbra, 3004-516, Portugal}
\affiliation[14]{
	Department of Physics and Astronomy, Texas A\&M University, College Station, TX 77843-4242, USA}
\affiliation[15]{
	Donostia International Physics Center (DIPC), Paseo Manuel Lardizabal, 4, Donostia-San Sebastian, E-20018, Spain}
\affiliation[16]{
	Escola Polit\`ecnica Superior, Universitat de Girona, Av.~Montilivi, s/n, Girona, E-17071, Spain}
\affiliation[17]{
	Departamento de F\'isica Te\'orica, Universidad Aut\'onoma de Madrid, Campus de Cantoblanco, Madrid, E-28049, Spain}
\affiliation[18]{
	Instituto de F\'isica Corpuscular (IFIC), CSIC \& Universitat de Val\`encia, Calle Catedr\'atico Jos\'e Beltr\'an, 2, Paterna, E-46980, Spain}
\affiliation[19]{
	Pacific Northwest National Laboratory (PNNL), Richland, WA 99352, USA}
\affiliation[20]{
	Instituto Gallego de F\'isica de Altas Energ\'ias, Univ.\ de Santiago de Compostela, Campus sur, R\'ua Xos\'e Mar\'ia Su\'arez N\'u\~nez, s/n, Santiago de Compostela, E-15782, Spain}
\affiliation[21]{
	Instituto de Instrumentaci\'on para Imagen Molecular (I3M), Centro Mixto CSIC - Universitat Polit\`ecnica de Val\`encia, Camino de Vera s/n, Valencia, E-46022, Spain}
\affiliation[22]{
	Laboratorio de F\'isica Nuclear y Astropart\'iculas, Universidad de Zaragoza, Calle Pedro Cerbuna, 12, Zaragoza, E-50009, Spain}
\affiliation[23]{
	Laboratorio Subterr\'{a}neo de Canfranc, Paseo de los Ayerbe s/n, Canfranc Estaci\'{o}n, Huesca, E-22880, Spain}

\abstract{Excellent energy resolution is one of the primary advantages of electroluminescent high-pressure xenon TPCs.  These detectors are promising tools in searching for rare physics events, such as neutrinoless double-beta decay (\bbonu), which require precise energy measurements.  Using the \NEW\ detector, developed by the NEXT (Neutrino Experiment with a Xenon TPC) collaboration, we show for the first time that an energy resolution of 1\% FWHM can be achieved at 2.6 MeV, establishing the present technology as the one with the best energy resolution of all xenon detectors for \bbonu\ searches.}


\collaboration{\includegraphics[height=9mm]{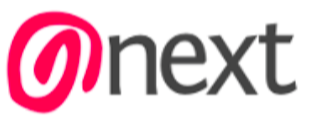}\\[6pt]
	NEXT collaboration}

\maketitle
\flushbottom

\section{Introduction}
Searches for neutrinoless double beta decay (\bbonu), the observation of which would imply total lepton number violation and would show that neutrinos are Majorana particles \cite{Schechter:1981bd, GomezCadenas:2013ue, Avignone_2008, GomezCadenas:2012fe}, require excellent energy resolution to eliminate background events that occur at energies similar to the Q-value of the decay (\Qbb).  The NEXT (Neutrino Experiment with a Xenon TPC) collaboration \cite{Alvarez:2011my, Alvarez:2012haa, Gomez-Cadenas:2013lta, Martin-Albo:2015rhw} intends to search for \bbonu\ using of order 100 kg of xenon enriched to 90\% in the candidate isotope $^{136}$Xe (\Qbb~=~2457.8 keV).  In recent years, NEXT has developed and operated several gaseous xenon TPCs, including kg-scale detectors at Lawrence Berkeley National Lab (LBNL) and Instituto de F\'{i}sica Corpuscular (IFIC) \cite{Alvarez:2012hh, Alvarez:2012xda} and more recently the 5 kg-scale \NEW\footnote{Named after our late mentor and friend Prof. James White.} at the Canfranc Underground Laboratory (LSC) in the Spanish Pyrenees \cite{Monrabal:2018xlr}.

Previous analyses \cite{Renner:2018csth} of the \NEW\ energy resolution using gamma rays from \CS\ and \THO\ sources showed an extrapolated 1\% FWHM resolution at \Qbb.  The relatively low pressure (\NewSevenBarPressureRunII) at which those data were taken meant that electron tracks of events with energy near \Qbb\ were not easily contained in the detector.  Low statistics at the photopeak limited the highest energy at which a detailed analysis of energy resolution was performed to 1.6 MeV.  More data has since been taken at a higher pressure (10.3 bar), and the results are reported in the present study.  The experimental setup, similar to that of the previous study \cite{Renner:2018csth}, is reviewed in section \ref{sec.setup}, and the analysis and obtained energy resolution are presented in section \ref{sec.resolution}.

\section{Experimental setup}\label{sec.setup}
\subsection{The \NEW\ electroluminescent TPC}

\begin{figure}[tbh]
\begin{minipage}[c]{0.62\textwidth}
	\includegraphics[width= 1.0\textwidth]{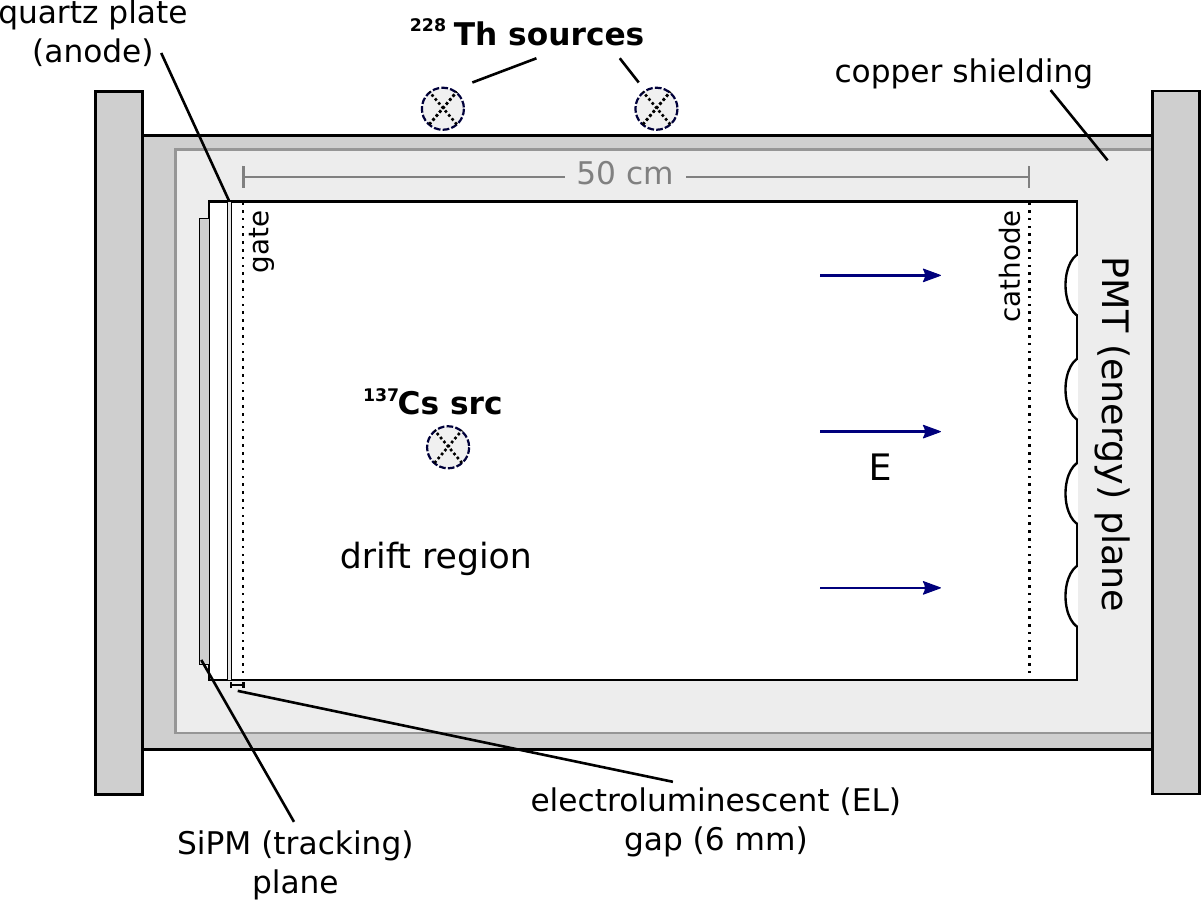}
\end{minipage}\hfill
\begin{minipage}[c]{0.32\textwidth}
\resizebox{1.0\textwidth}{!}{%
\begin{tabular}{|c|c|}
	\hline
	TPC parameter & Value \\
	\hline
	Pressure & 10.3 bar\,* \\
	$E/p$ & 1.3 kV cm$^{-1}$ bar$^{-1}$ \\
	Drift field & 415 V cm$^{-1}$ \\
	$V_{\mathrm{cathode}}$ & -30 kV \\
	$V_{\mathrm{gate}}$ & -7.9 kV \\
	Length & \NewTpcLength \\
	Diameter &  \NewTpcDiameter \\
	EL gap & \NewTpcELGap \\
	Drift length & \NewTpcDriftLength \\
	Fiducial mass & 3.3 kg \\
	\hline\hline
\end{tabular}}
\footnotesize{*\,The actual measured pressures for each run varied between 10.27-10.32 bar.}
\end{minipage}
\caption{Experimental summary.  (Left) Schematic of the main components of the \NEW\ TPC and locations of the calibration sources (not drawn to scale).  \CS\ and \TO\ sources were placed in the lateral and top entrance ports of the pressure vessel, respectively, and a second \TO\ source (the leftmost of the two) was placed directly on top of the vessel.  (Right) Operational parameters used in the present study.}
\label{fig.config}
\end{figure}

The experimental setup is similar to that of the preceding study \cite{Renner:2018csth} and is summarized here.  The detector \NEW\ is an electroluminescent (EL) time projection chamber (TPC) filled with xenon gas and equipped with photosensors to detect the UV light emitted in interactions occurring within the active volume.  Charged particles deposit energy within the drift region, producing a track of ionized and excited xenon atoms.  The UV light emitted in the relaxation of the excited xenon atoms, called primary scintillation or S1, is detected immediately and the ionized electrons are drifted toward a readout plane consisting of a narrow region of high electric field, the EL gap.  In passing through the EL gap, the electrons are accelerated to energies high enough to further excite, but not ionize, the atoms of the xenon gas, leading to the production of an amount of secondary scintillation photons (S2) proportional to the number of electrons traversing the gap.  This amplification process, electroluminescence, allows for gains on the order of 1000 photons per electron with significantly lower fluctuations than avalanche gain.  In addition, the time elapsed between the observation of S1 and the arrival of S2 can be used to determine the axial ($z$) coordinate at which the interaction took place.

In \NEW\ (see \Fig\ \ref{fig.config} and also \cite{Monrabal:2018xlr}), the primary (S1) and secondary (S2) scintillation are detected by an array of \NewNumberOfPMT\ \NewTypePMT\ photomultiplier tubes (PMTs), called the ``energy plane'' placed \NewCathodeToPMTs\ from a transparent wire mesh cathode held at negative high voltage.  An electric field is established in the drift region defined by the cathode and another transparent mesh (the gate) located about 53 cm away.  The EL region is defined by the mesh and a grounded quartz plate coated with indium tin oxide (ITO), placed \NewTpcELGap\ behind it.  A grid (\NewSipmPitch\ pitch) of \NewNumberOfSiPM\ SensL series-C silicon photomultipliers (SiPMs) is located behind the EL gap and measures the S2 scintillation, providing precise information on where the EL light was produced in $(x,y)$.  The active volume is shielded by an \NewBarrelICS\ thick ultra-pure inner copper shell, and the sensor planes are mounted on pure copper plates of thickness \NewPlatesICS.  The sensor planes and active volume are enclosed in a pressure vessel constructed from the titanium-stabilized stainless steel alloy \NewPressureVesselMaterial.  The vessel sits on top of a seismic table; and a lead shield that can be mechanically opened and closed surrounds the vessel.  The vessel is connected to a gas system through which the xenon gas is continuously purified via the use of a hot getter.  The entire experimental area, including gas system, electronics, pressure vessel, and seismic table, are stationed on an elevated tramex platform in the Laboratorio Subterr\'{a}neo de Canfranc (LSC) in the Spanish Pyrenees.

\begin{table}
	\begin{center}
		\caption{Summary of data analyzed in this study.}\label{tbl.runs}
		\begin{tabular}{ccrrrr}
			& & Avg& Triggers & Triggers & Avg\\
			Run \# & Duration & Rate & (low-energy) & (high-energy) & Lifetime (\textmu s)\\
			\hline
			6346 & 25.0 h & 42 Hz & 3 485 555 & 313 761 & 3977\\
			6347 & 23.6 h & 41 Hz & 3 250 612 & 304 948 & 4190\\
			6348 & 23.5 h & 41 Hz & 3 210 597 & 307 397 & 4297\\
			6349 & 23.8 h & 41 Hz & 3 248 563 & 311 204 & 4261\\
			6351 & 23.9 h & 41 Hz & 3 260 929 & 311 951 & 4008\\
			6352 & 24.6 h & 41 Hz & 3 345 650 & 321 545 & 3908\\
			6365 & 24.4 h & 41 Hz & 3 300 055 & 318 662 & 3344\\
			6482 & 26.7 h & 41 Hz & 3 257 113 & 739 668 & 3527\\
			6483 & 24.7 h & 41 Hz & 3 006 991 & 684 718 & 3579\\
			6484 & 24.4 h & 41 Hz & 2 959 826 & 681 687 & 3586\\
			6485 & 20.3 h & 41 Hz & 2 453 528 & 566 984 & 3597\\
		\end{tabular}
	\end{center}
\end{table}

\subsection{Run configuration}
\begin{figure}[htb]
	\centering
	\includegraphics[width= 1.0\textwidth]{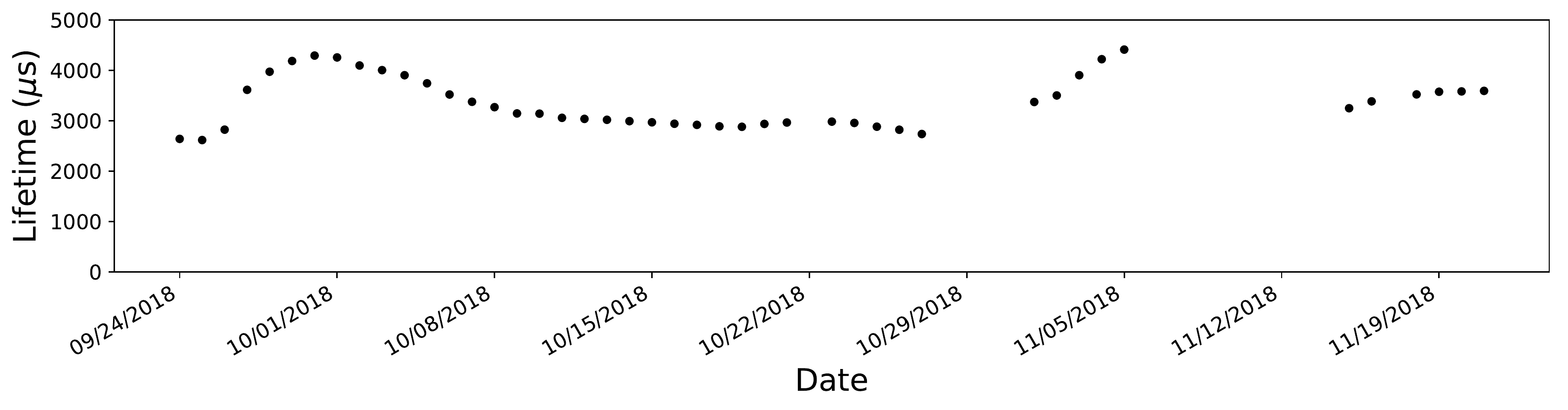}
	\caption{The average electron lifetime over the course of the analyzed runs determined using \Kr{83m} events.  The errors on the measurements are smaller than the size of the points.  Though consistently above 2 ms, the electron lifetime was unstable and therefore was monitored and the corresponding corrections determined for each run.}\label{fig.lifetime}
\end{figure}

As the goal of the present analysis was a detailed study of energy resolution, calibration sources were employed to yield energy peaks over a range of energies from several tens of keV up to and including \Qbb.  \Kr{83m} was injected into the xenon gas, providing a uniform distribution of \KrEnergy\ point-like energy depositions used to map out the geometric variations in the sensor responses and electron lifetime of the detector \cite{Martinez-Lema:2018ibw}.  \CS\ and \TO\ calibration sources were also placed as shown in \Fig\ \ref{fig.config}.  The \CS\ source provided \CsGammaEnergy\ gamma rays, and \TO\ decays to \TL\, which provides gamma rays of energy 2614.5 keV. In this study we focus on the energy peaks produced by interactions of these \CS\ and \TL\ gamma rays, and also the double-escape peak resulting from $e^{+}e^{-}$ pair production interactions of the \TL\ gamma ray in which the two 511 keV gamma rays escape.  For the present analysis, the acquisition trigger was split into a lower-energy trigger seeking the \Kr{83m} events and a high-energy trigger aimed at capturing events with energy above approximately 150 keV.  A summary of the datasets analyzed is given in Table \ref{tbl.runs}.  For each run, the low-energy triggers were used to compute the lifetime and geometric correction maps used to correct the events acquired with the high-energy trigger.  The average electron lifetime determined over the course of the analyzed runs is also shown in \Fig\ \ref{fig.lifetime}.  Variations in the average lifetime were observed across runs, most likely due to outgassing of internal components, and for this reason corrections for each run needed to be determined using data taken over the same period of time.

\section{Energy resolution}\label{sec.resolution}
\begin{figure}[htb]
	\centering
	\includegraphics[width= 1.0\textwidth]{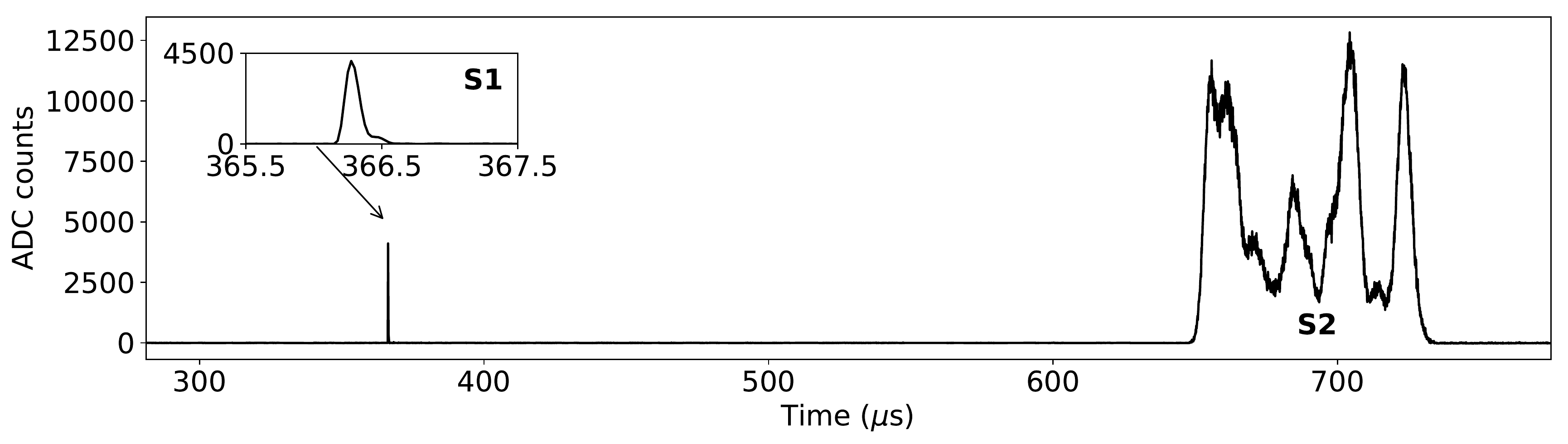}
	\caption{The acquired waveform, summed over all PMTs, for an event in the \Tl{208} photopeak.  Note that this particular event was identified to contain a single continuous track, as evident partially in the existence of a single long S2 pulse.}\label{fig.waveform}
\end{figure}

The signals from the SiPMs and PMTs were digitized in samples of width 1 \textmu s and 25 ns respectively. Individual pulses in the energy plane waveform (summed over all PMTs, see \Fig\ \ref{fig.waveform}) were selected and classified as either S1 or S2.  Events with a single identified S1 were selected, and the S2 peaks were divided into ``slices'' of width 2 \textmu s.

\begin{figure}[htb]
	\centering
	\includegraphics[width= 1.0\textwidth]{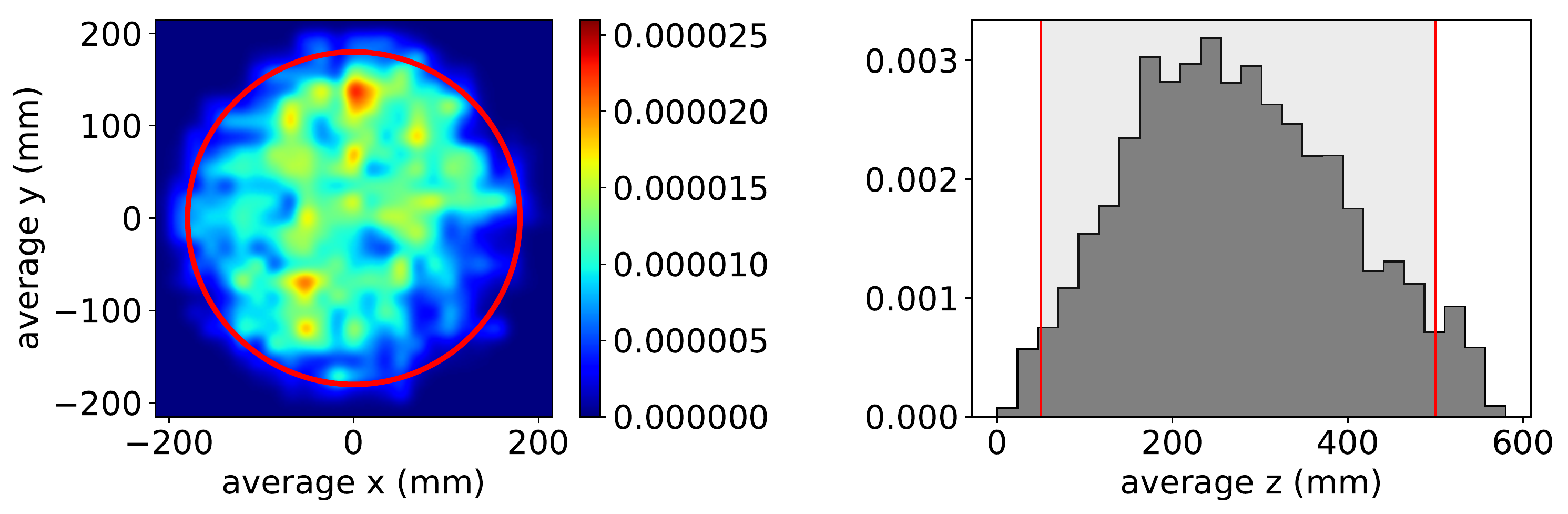}
	\caption{Normalized distributions in $x$-$y$ (left) and $z$ (right) of observed energy depositions for events in the 2615 keV \Tl{208} photopeak (data from run 6485).  The solid red lines show the fiducial cuts employed in this study which encompass nearly the entirety of the active volume.  Note that the fiducial cuts are placed on all reconstructed energy depositions, rejecting an event if one or more depositions occurred outside the cut.}\label{fig.fiducial}
\end{figure}

The pattern of light detected by the SiPMs of the tracking plane during the 2 \textmu s interval of the slice was used to reconstruct the $(x,y)$ location of the EL production, as done in \cite{Martinez-Lema:2018ibw}, except multiple reconstructed positions sharing the energy $E$ of a single slice were possible in order to allow for reconstruction of tracks that extend over greater distances (several cm or more) in the $(x,y)$ plane within a single slice.  The time elapsed since the S1 pulse was used to determine the $z$ coordinate of each slice, and the energies $E$ of the reconstructed depositions $(x,y,z,E)$ were then multiplied by two correction factors: one accounting for the geometrical $(x,y)$ dependence of the light collection over the EL plane, and another accounting for losses due to the finite electron lifetime caused by attachment to impurities.  This second factor depended on the drift length ($z$-coordinate) and the location in the EL plane $(x,y)$, as the electron lifetime also varied in $(x,y)$.  Once fully reconstructed, fiducial cuts were made on each event as detailed in \Fig\ \ref{fig.fiducial}.

\begin{figure}[htb]
	\centering
	\includegraphics[width= 0.48\textwidth]{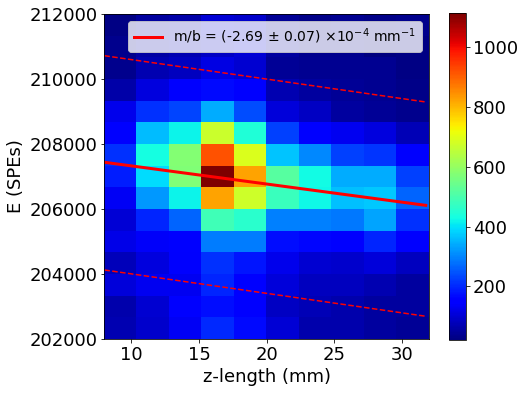}
	\includegraphics[width= 0.48\textwidth]{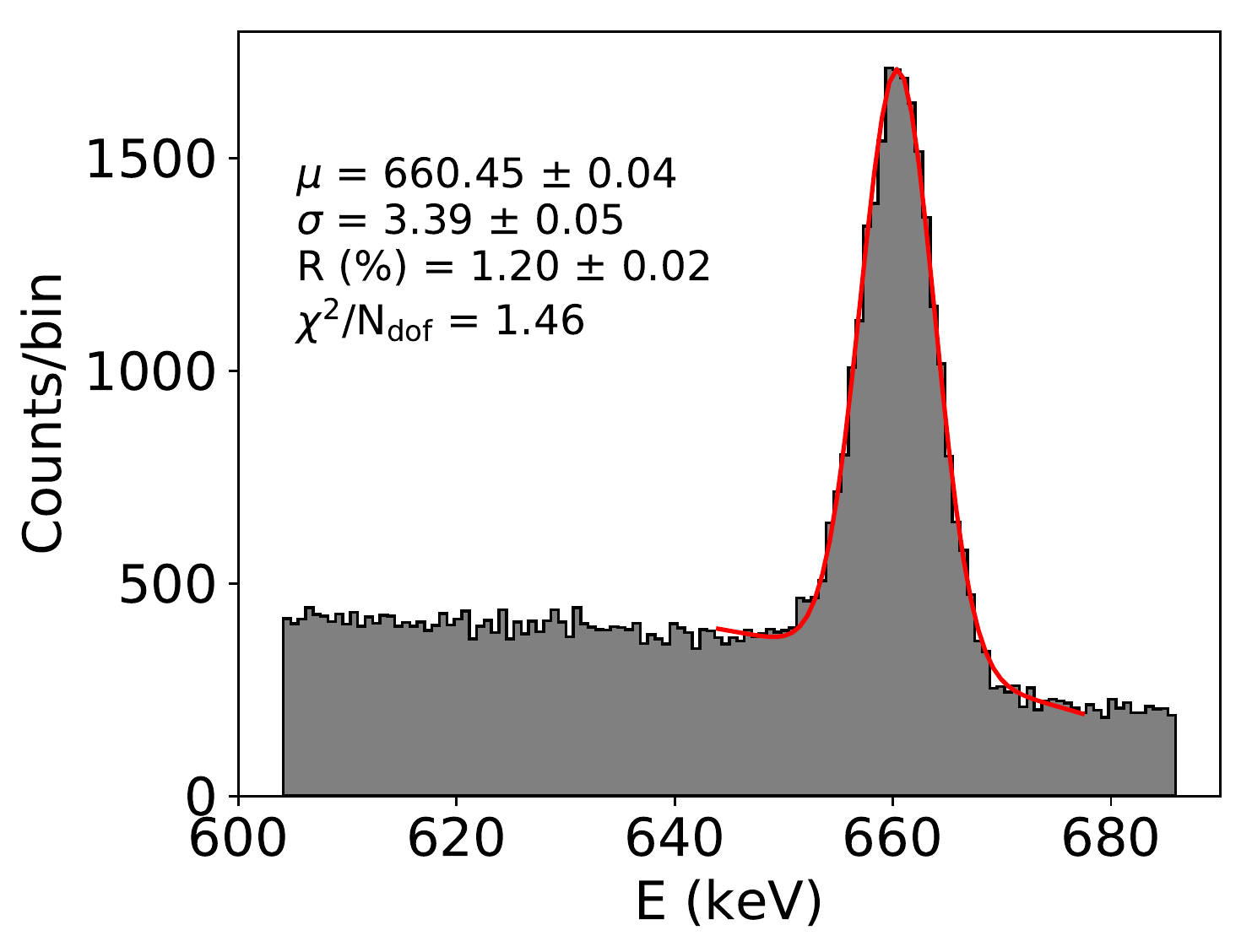}
	\includegraphics[width= 0.48\textwidth]{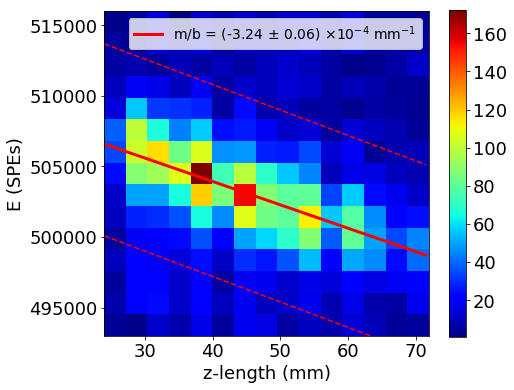}
	\includegraphics[width= 0.48\textwidth]{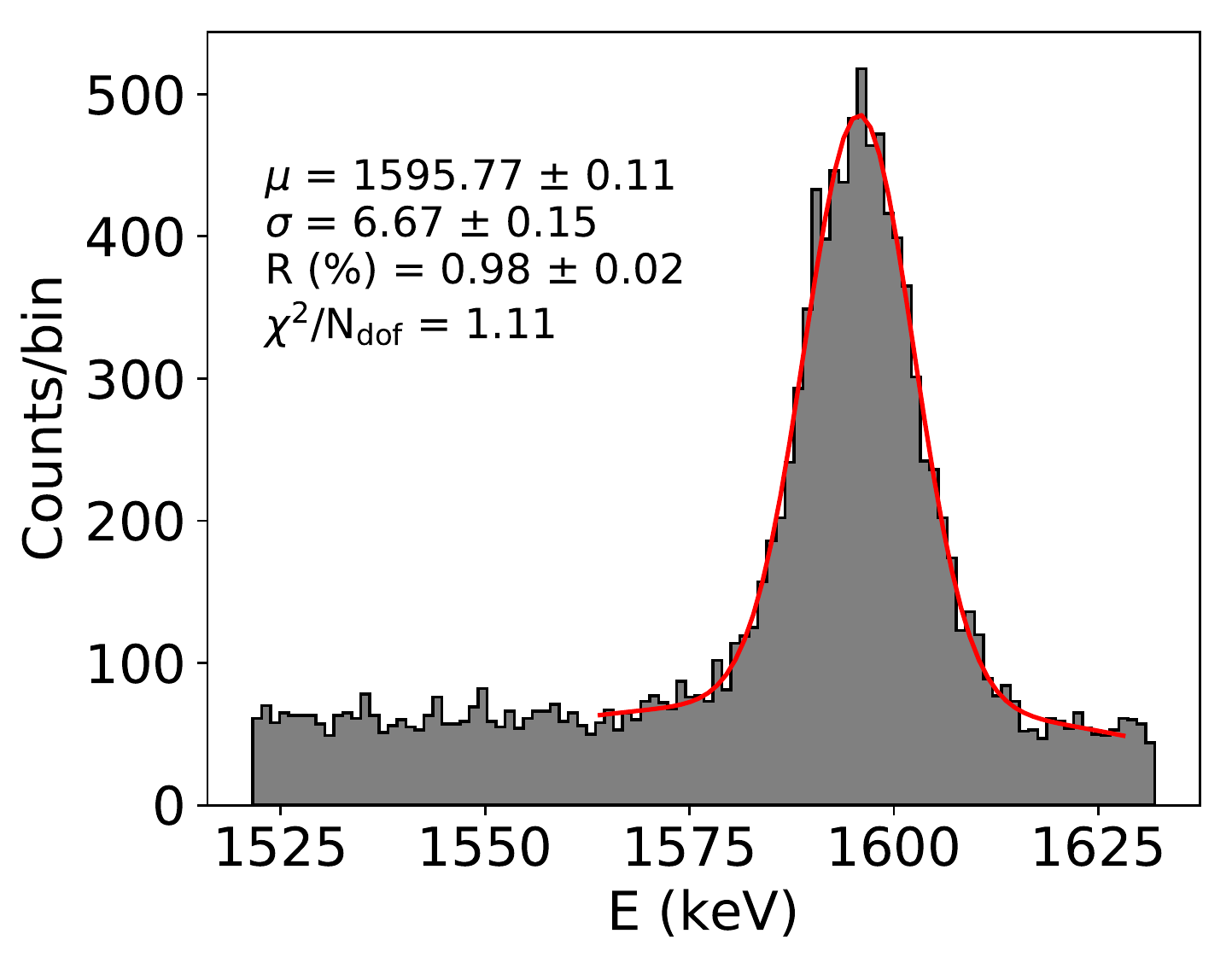}
	\includegraphics[width= 0.48\textwidth]{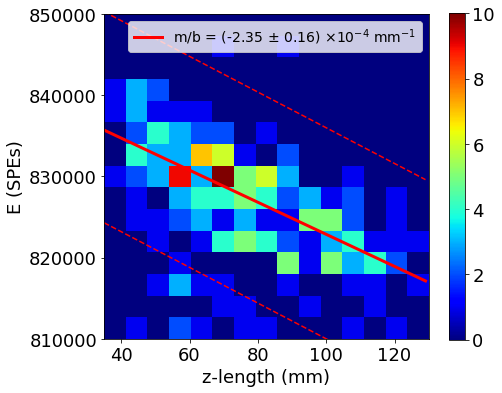}
	\includegraphics[width= 0.48\textwidth]{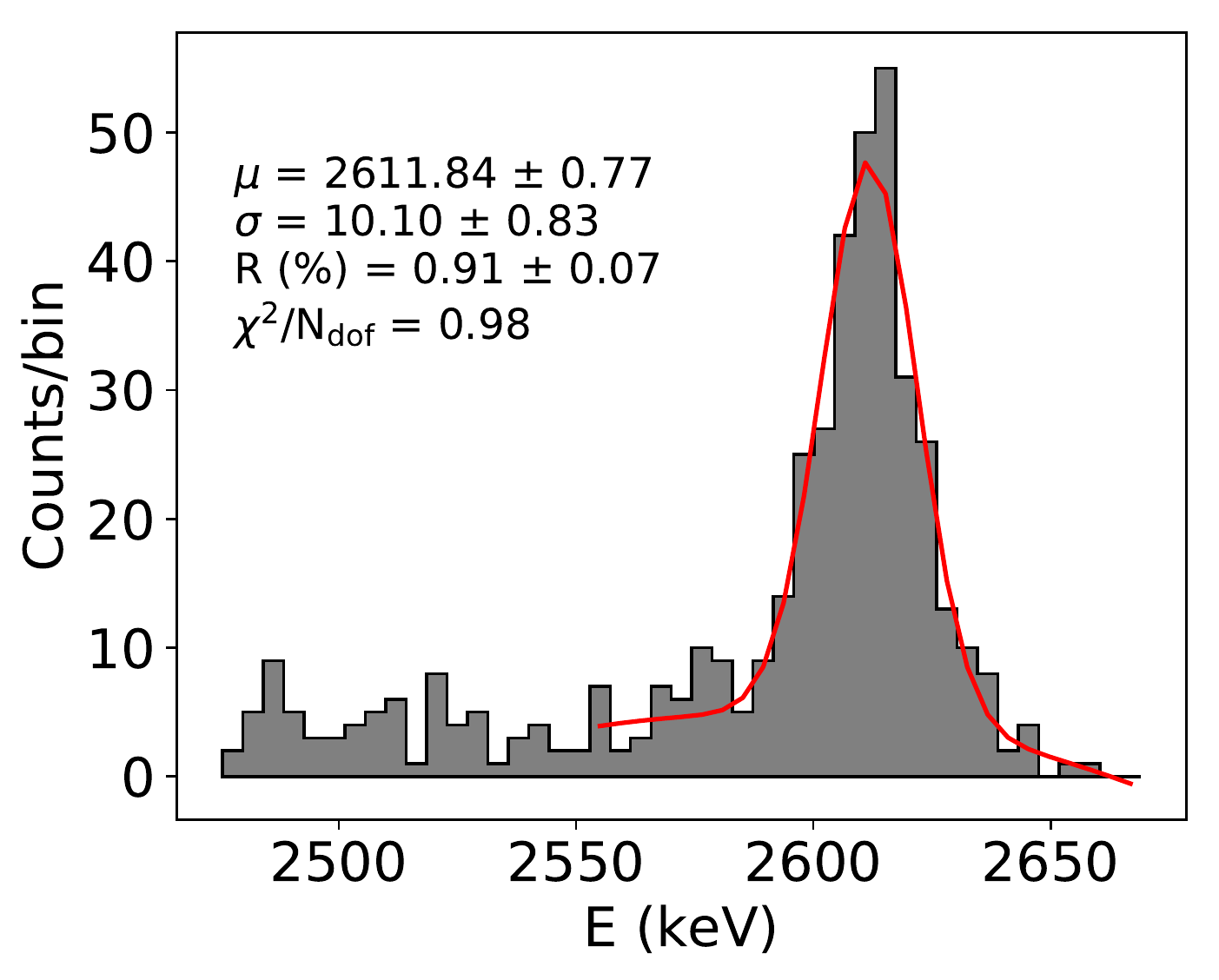}
	\caption{Fits to the dependence of energy on track length in the axial dimension (left), and the resulting energy spectra of three energy peaks (nominally at 662 keV, 1592 keV, and 2615 keV) after application of all corrections, including a linear correction to the energy (equation \ref{eqn.ecorrection}) corresponding to the average value of $(m/b) = 2.76\times 10^{-4}$ obtained from the 3 fits (right).}\label{fig.resolution}
\end{figure}

A final correction was applied for an empirically observed dependence on the track orientation.  The origin of this dependence, whereby the measured energy of an event decreases with increasing axial ($z$) extent of the track, is still under study.  However, we find it can be effectively corrected, as follows.  The z-extent $\Delta z$ is defined as the difference between the maximum and minimum z-coordinates of all reconstructed slices in the event.  The effect is shown in \Fig\ \ref{fig.resolution} along with the resolution obtained for each of the three peaks (662 keV, 1592 keV, and 2615 keV) after correcting for the effect using the average of the normalized slopes determined by a linear fit to each distribution,

\begin{equation}\label{eqn.ecorrection}
E_{\mathrm{corrected}} = \frac{E}{1 - (m/b)\Delta z},
\end{equation}

\noindent where $m$ and $b$ are the slope and intercept of the linear fit for $\Delta z$ in mm.  Note that the linear fits were performed on the events between the dashed lines.  Reasonable variations on the positioning of these lines (repositioning them vertically by several thousand SPEs without visibly cutting into the dense areas of the 2D distributions) gave an error of approximately $0.2 \times 10^{-4}$ for each computed $(m/b)$ in addition to the statistical errors shown on the distributions in \Fig\ \ref{fig.resolution} (left).   In determining $(m/b)$ and in the subsequent determination of energy resolution, all events were required to have z-lengths in the ranges shown on the x-axes of the 2D distributions.  Furthermore, in order to avoid complications in the spectrum caused by interactions producing isolated secondary depositions such as Compton scattering, bremsstrahlung, and the emission of characteristic X-rays, all events were required to have been reconstructed as single continuous tracks.  To demonstrate the validity of the correction over time, data from runs 6346, 6347, 6348, 6349, and 6351 were used to determine $(m/b)$, and the remaining data, runs 6352, 6365, 6482, 6483, 6484, and 6485, were used to evaluate the energy resolution (see Table \ref{tbl.runs}).

Each peak was fitted to the sum of a Gaussian and a 2nd-order polynomial to account for the surrounding distribution of background events, and the resolution was computed using the width of the Gaussian.  The obtained resolutions are: $1.20 \pm 0.02$\% FWHM at 662 keV; $0.98 \pm 0.03$\% FWHM at 1592 keV; and $0.91 \pm 0.12$\% FWHM at 2615 keV.  The total errors are estimated in each case based on the statistical errors of the fits (shown on the histograms in \Fig\ \ref{fig.resolution}) and systematic effects including variations in the range of events included in the fit and (systematic) errors in the correction for the axial length effect.  The energy conversion from detected photoelectrons to keV was determined (after application of all corrections) using a quadratic fit to the means of the three peaks of interest (662 keV, 1592 keV, 2615 keV) and the 29.7 keV K-$\alpha$ xenon X-ray peak.  The X-rays had energies too low to be triggered on as individual events, but their energies were visible by examining the spectrum of isolated energy depositions within all events, which included small depositions due to xenon X-rays that traveled away from the main track before interacting.  

\begin{figure}[htb]
	\centering
	\includegraphics[width= 1.0\textwidth]{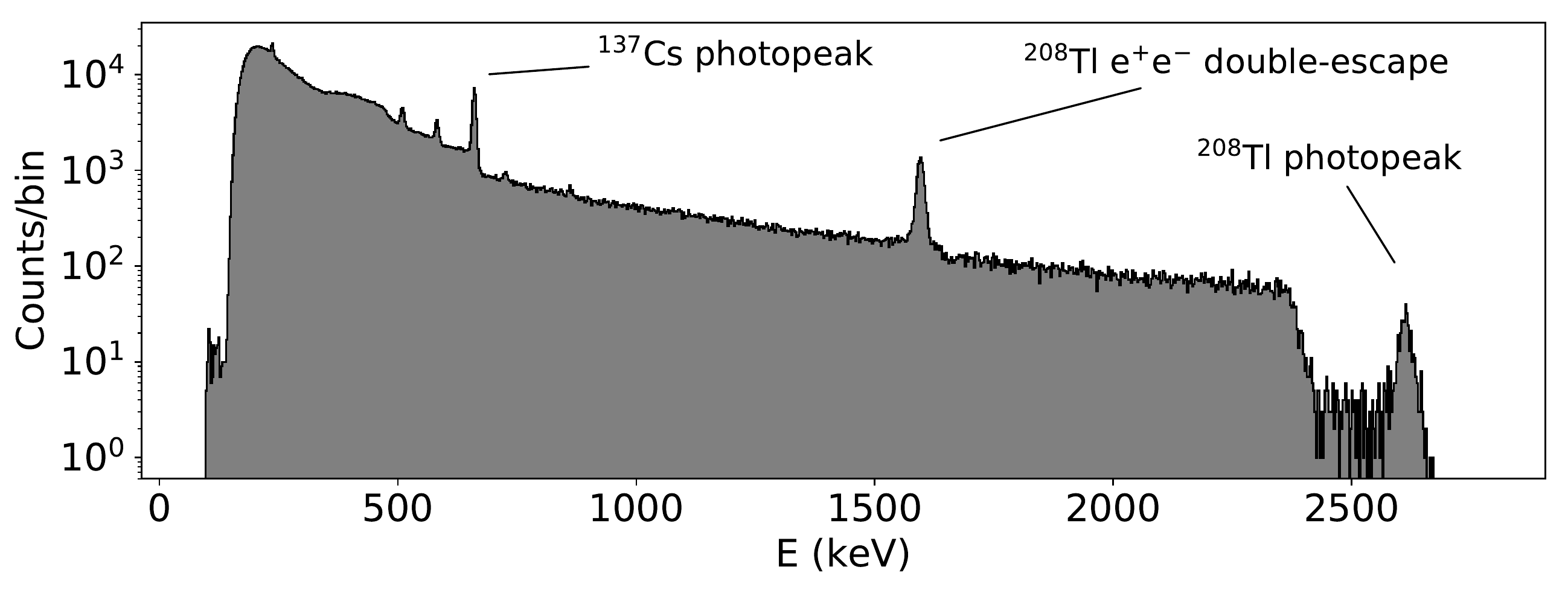}
	\caption{The full energy spectrum for events with energies greater than approximately 150 keV.  Corrections for electron lifetime and geometrical effects were applied to all events, as well as a correction for the described axial length effect corresponding to $(m/b) = 2.76\times 10^{-4}$.  In addition to the three lines examined in detail in this study, lines are also present due to other gamma rays from the $^{228}$Th decay chain: at 238 keV (from $^{212}$Pb $\rightarrow$ $^{212}$Bi decay), 511 keV (e$^{+}$e$^{-}$ annihilation, with some contribution from $^{208}$Tl $\rightarrow$ $^{208}$Pb decay), 583 keV ($^{208}$Tl $\rightarrow$ $^{208}$Pb decay), 727 keV ($^{212}$Bi $\rightarrow$ $^{212}$Po decay), and 860 keV ($^{208}$Tl $\rightarrow$ $^{208}$Pb decay).}\label{fig.spectrum}
\end{figure}

\begin{figure}[htb]
	\centering
	\includegraphics[width= 1.0\textwidth]{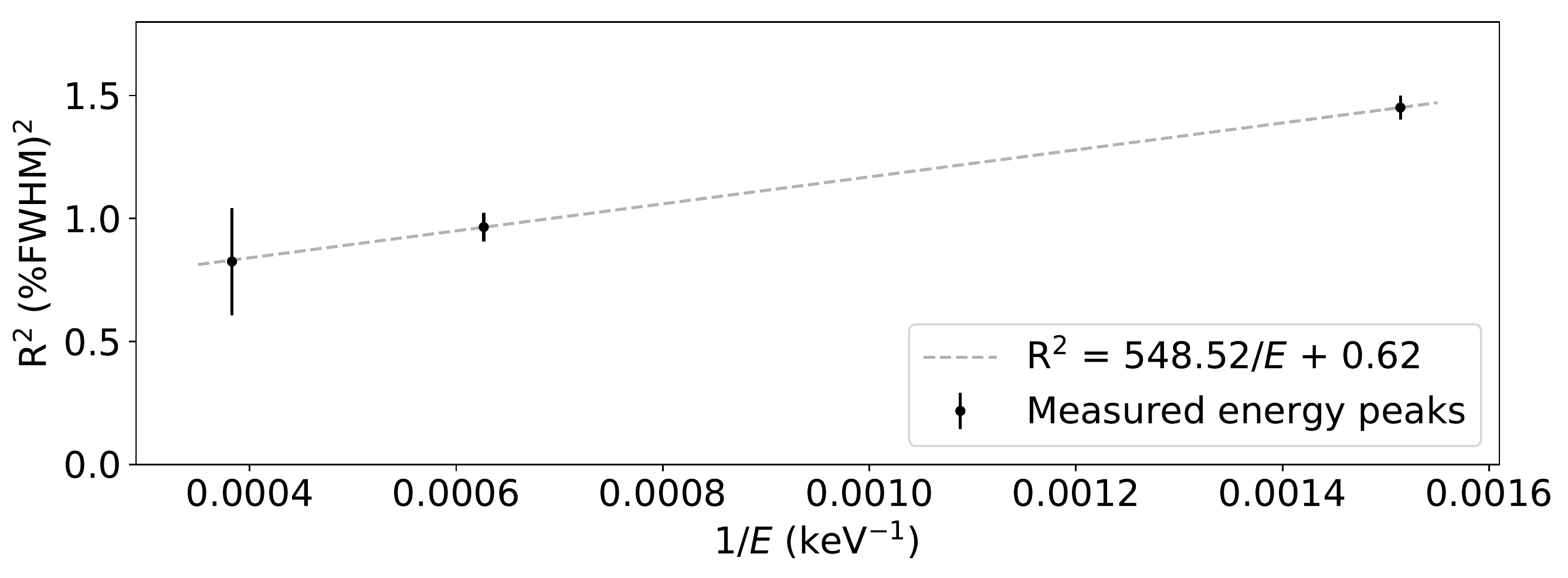}
	\caption{Squared energy resolution plotted against $1/E$.  The measured points were fit to the functional form $R^{2} = a/E + c$, with $a = 548.52 \pm 82.15$ \%FWHM$^{2}\,\cdot\,$keV and $c = 0.62 \pm 0.10$ \%FWHM$^{2}$.}\label{fig.Rsummary}
\end{figure}

The energy spectrum of high-energy triggers in the full active volume is shown in \Fig\ \ref{fig.spectrum} after applying all corrections described in the present section.  Unlike in the previous study \cite{Renner:2018csth}, the \Tl{208}\ photopeak at 2615 keV (near \Qbb) is clearly resolved.  The squared resolution is shown vs. the inverse energy in \Fig\ \ref{fig.Rsummary} for the three energy peaks studied and fit to a line, $R^{2} = a/E + c$, where $a = 548.52 \pm 82.15$ \%FWHM$^{2}\,\cdot\,$keV and $c = 0.62 \pm 0.10$ \%FWHM$^{2}$. The presence of a constant term in the resolution shows that detector-specific systematic effects have not been completely eliminated in our analysis, and there is still room for further improvement.  Nevertheless, these results demonstrate that excellent energy resolution is obtainable throughout the entire fiducial volume once correction for the axial length effect is made.  

Unlike corrections for electron lifetime, the correction for the axial length effect was relatively stable throughout the time (about 8 weeks) over which the data presented in this study was taken.  This effect is thought to be a result of some internal nonlinearity in light production, possibly due to alterations of the EL field during track readout in a ``charging-up'' effect as electrons cross the EL gap, or due to electron loss in the EL gap from attachment to impurities produced by photoionizing the wavelength shifting material deposited on the surface behind the EL region.  Other potential explanations such as PMT saturation, electron recombination with the ions of the original track, variations in electron lifetime throughout the detector, and emission of light by the SiPMs, have already been discredited.  Further details are given in appendix \ref{app.axialeffect}.

\section{Summary}
\label{sec.conclu}

Energy resolution in the \NEW\ TPC has been further studied, and a resolution better than 1\% FWHM is shown to be obtainable at 2615 keV, as predicted in the preceding study \cite{Renner:2018csth}.  This resolution was obtained over nearly the entire active volume, demonstrating the effectiveness of the continuous \Kr{83m}-based calibration procedure implemented to correct for geometric and lifetime effects, and improved slightly with more restrictive fiducial cuts.  Further study is required to understand the observed ``axial length effect'' in which the measured energy of extended tracks decreases with increasing track length in the axial (drift) direction.  However, as high-pressure xenon TPCs provide detailed energy and topological information for each event, such effects can be remedied through careful calibration, and the outstanding resolution obtained highlights the strong potential of this detector technology to host a sensitive $0\nu\beta\beta$ search in which good energy resolution is essential.

\acknowledgments
The NEXT Collaboration acknowledges support from the following agencies and institutions: the European Research Council (ERC) under the Advanced Grant 339787-NEXT; the European Union's Framework Programme for Research and Innovation Horizon 2020 (2014-2020) under the Marie Sk\l{}odowska-Curie Grant Agreements No. 674896, 690575 and 740055; the Ministerio de Econom\'ia y Competitividad and the Ministerio de Ciencia, Innovaci\'on y Universidades of Spain under grants FIS2014-53371-C04, RTI2018-095979, the Severo Ochoa Program SEV-2014-0398 and the Mar\'ia de Maetzu Program MDM-2016-0692; the GVA of Spain under grants PROMETEO/2016/120 and SEJI/2017/011; the Portuguese FCT under project PTDC/FIS-NUC/2525/2014, under project UID/FIS/04559/2013 to fund the activities of LIBPhys, and under grants PD/BD/105921/2014, SFRH/BPD/109180/2015 and SFRH/BPD/76842/2011; the U.S.\ Department of Energy under contracts number DE-AC02-06CH11357 (Argonne National Laboratory), DE-AC02-07CH11359 (Fermi National Accelerator Laboratory), DE-FG02-13ER42020 (Texas A\&M) and DE-SC0019223 / DE-SC0019054 (University of Texas at Arlington); and the University of Texas at Arlington. DGD acknowledges Ramon y Cajal program (Spain) under contract number RYC-2015-18820. We also warmly acknowledge the Laboratori Nazionali del Gran Sasso (LNGS) and the Dark Side collaboration for their help with TPB coating of various parts of the NEXT-White TPC. Finally, we are grateful to the Laboratorio Subterr\'aneo de Canfranc for hosting and supporting the NEXT experiment.

\bibliographystyle{pool/JHEP}
\bibliography{pool/NextRefs}

\appendix
\section{The axial length effect}\label{app.axialeffect}
The origin of the axial length effect is still under detailed study. However, a number of possible origins have already been discarded:

\begin{enumerate}
	\item[\textbullet]\textbf{PMT saturation / baseline shift:} due to the AC-coupled PMT readout scheme used in \NEW\ \cite{Monrabal:2018xlr}, all PMT waveforms must be passed through a deconvolution algorithm to remove distortions introduced by high-pass filtering before physics analysis.  It was found that if the response of a PMT saturates, the deconvolution may lead to a shifted baseline which could lead to an error in the signal integration (energy) dependent on the length of integration in time ($z$).  However, the effect was found to persist even after lowering PMT gains, ensuring no saturation, and it was confirmed that any shift in baseline present after the deconvolution was not significant enough to account for the observed effect.
	\item[\textbullet]\textbf{Recombination:} as the electrons are drifted in the $z$-dimension towards the EL plane, it was proposed that tracks extended in this dimension present a greater opportunity for drifting electrons to encounter neighboring ions and recombine.  Since these electrons would not arrive at the EL plane and produce light, this would lead to a lower energy measurement.  However, basic simulations concluded that the recombination capture radius would need to be on the order of several tens of \textmu m to explain the effect, an unphysically large sphere of influence for a single ion.  In addition, electron-ion recombination would lead to scintillation light that should be observable during a time interval beginning after primary scintillation and ending after an amount of time required to drift the electrons over the entire track length in $z$.  For \Tl{208} photopeak events (see \Fig\ \ref{fig.resolution}, bottom), this would be about 120 \textmu s, and integrating over this interval after the arrival of S1 for many such events, no evidence of the expected light was observed.
	\item[\textbullet]\textbf{Variations in electron lifetime:} as the measured electron lifetime in \NEW\ is known to vary with location in the detector, there has been concern that small errors in the computation of the lifetime were giving rise to the observed effect when applied over long tracks.  However, even after correcting Cs-photopeak events using a single average position (assuming pointlike tracks), the effect could still be observed by making a tight cut on average radius (effectively eliminating the error due to response variations in the $xy$-plane by considering only events that did not require significant $xy$ correction).
	\item[\textbullet]\textbf{Light emitted from the SiPMs:} the effect is also seen in the integrated charge of the SiPMs, and in fact is more dramatic (the normalized slopes $m/b$ analogous to those shown in \Fig\ \ref{fig.resolution} are greater in magnitude).  Therefore it was proposed that the SiPMs may be emitting additional light in a nonlinear manner during the production of EL.  However, even after turning off the SiPM plane and using only information from the PMT plane for a less-precise $xy$ reconstruction, the effect was still observed.
\end{enumerate}

\noindent Several explanations for the effect have not yet been investigated in detail:

\begin{enumerate}
	\item[\textbullet]\textbf{``Charge-up'' effect at the EL plane:} an electron crossing the EL gap may, at least locally, alter the electric field seen by the next electron crossing the gap for some amount of time.  If this were to make the average gain somewhat dependent on track orientation - whether the electrons cross the gap more in ``series'' (more extended in $z$) or in ``parallel'' (more extended in $xy$) - this could give rise to the observed effect. 
	\item[\textbullet]\textbf{Attachment to ionized impurities in the EL gap:} The wavelength shifter tetraphenyl butadiene (TPB) is deposited on several components in \NEW\, including the quartz plate just behind the EL region, to shift the VUV scintillation produced by xenon to visible light that can be detected by the photosensors (the SiPMs are not VUV sensitive, and the PMTs are placed inside enclosures behind sapphire windows, which do not transmit VUV light, to shield them from the high pressure environment inside the detector).  If the photons produced in electroluminescence are capable of photoionizing the TPB, the resulting ions would be drifted across the EL region, possibly capturing some of the electrons that arrived at later times before completely traversing the EL gap and thereby reducing the observed energy of the event.
\end{enumerate}

 The observed effect could also be a result of a nonlinearity in the light production process caused by some other internal component.  Further investigation in future runs with \NEW, possibly involving alterations of the internal hardware and/or running systematically at different EL gains, will be necessary to understand this effect.  Nevertheless, excellent resolution has been obtained due to the properties of high-pressure xenon TPCs, such as simultaneous energy and 3D position measurements, which allow for detailed calibration.

\end{document}